\documentclass[12pt]{iopart}

\expandafter\let\csname equation*\endcsname=\relax 
\expandafter\let\csname endequation*\endcsname=\relax 
\usepackage{amsmath,amssymb,amsthm} 
\usepackage{dsfont,graphicx,bm,caption,subcaption}
\usepackage{xcolor}

\newcommand{\p}{\partial}

\begin{document}

\title[Order-by-disorder in magnets with frustrated spin interactions]{Order-by-disorder in magnets with frustrated spin interactions - classical and large-$S$ limits via the spin functional integral}
\author{Peng Rao$^1$, Johannes Knolle$^{1,2,3}$}
\address{$^1$ Physics Department, Technical University of Munich, TUM School of Natural Sciences, 85748 Garching, Germany}
\address{$^2$ Munich Center for Quantum Science and Technology (MCQST), Schellingstr. 4, 80799 München, Germany}
\address{$^3$ Blackett Laboratory, Imperial College London, London SW7 2AZ, United Kingdom}
\ead{peng.rao@tum.de}

\begin{abstract}
	We investigate spin systems with extensive degeneracies in the classical ground states due to anisotropic frustrated spin interactions, where the degeneracy is not protected by symmetry. Using spin functional integration, we study the lifting of the degeneracies by fluctuations called order-by-disorder (ObD), and the associated gap in the spin-wave spectrum. It is shown that ObD corresponds to gradient-dependent anisotropic interactions of the pseudo-Goldstone modes, which vanish for a classical uniform spin configuration. Fluctuations generate a gradient-independent effective potential which determines the ground state and the pseudo-Goldstone gap. Furthermore, we recover previous predictions for the pseudo-Goldstone gap in type-I and II ObD with two-spin interactions in the large spin-$S$ limit or the classical small temperature limit, by computing the gap explicitly for the type-II cubic compass model and the type-I square compass model. We show that these two limits correspond to the one-loop approximation for the effective potential. We also discuss other types of order by disorder due to $m$-spin interactions where $m>2$.
\end{abstract}

Keywords: \noindent{\it Frustrated magnetism, Order by disorder, Pseudo-Goldstone mode \/} \\

\submitto{Journal of Physics: Condensed Matter}
\maketitle

\section{Introduction}
\label{sec:intro}
Order-by-disorder (ObD) describes spin systems in which the classical ground states are highly degenerate, the degeneracy being `accidental' since it is not related to the system symmetry and is removed by fluctuation corrections~\cite{Villain1980,Shender1982,Henley1989}. In such systems, the pseudo-Goldstone oscillations around the classical ground states are gapless to leading order, but they acquire a small gap once the fluctuations are taken into account. For this reason ObD has been proposed to explain the unexpected small gap in materials in particular in frustrated magnets ~\cite{Champion2003,Zhitomirsky2012,Savary2012}. ObD was also studied in model Hamiltonians on the square~\cite{chandra1990ising,Danu2016}, triangular~\cite{chubukov1992triangular}, Kagome~\cite{chubukov1992kagome,Reimers1993} and diamond lattices~\cite{Henley1994}, in the Kitaev honeycomb model~\cite{Baskaran2008} and on the pyrochlore lattice~\cite{Bramwell1994,McClarty2014}. Even though the large degeneracy arises as an artifact of the classical zero-temperature approximation, the pseudo-Goldstone gap determines the low-temperature properties of the system and is clearly experimentally observable. Therefore experimental confirmation of ObD in materials requires understanding the nature of the pseudo-Goldstone gap and its dependence on various physical parameters such as temperature.  

This problem was addressed in several theoretical works~\cite{Rau2018,Khatua2023}. In Ref.~\cite{Rau2018}, a general classification of ObD and the associated pseudo-Goldstone gap $\Delta$ has been established, in which, by considering interaction corrections in the Holstein-Primakoff approximation, the authors analyzed the algebraic structure of the magnon self-energy in the subspace of pseudo-Goldstone modes. It is found that there are two types of ObD distinguished by the form of their linear spin-wave spectrum: 
\begin{equation}\label{eq:ObD-dispersion}
	\text{type-I:}\ \omega \propto k,\ \text{type-II:}\ \omega \propto k^2.
\end{equation}
They give different dependence of $\Delta$ on the spin $S$ at zero temperature:
\begin{align}\label{eq:PG-gap-S}
	\text{type-I:}\ \Delta \sim \sqrt{S} , \ \text{type-II:}\ \Delta \sim O(S),
\end{align}
This result is followed up by Ref.~\cite{Khatua2023}, which showed by considering the type-I square-compass model that, for the two aforementioned cases, classical thermal fluctuations also produce a gap with distinct temperature dependence:
\begin{align}\label{eq:PG-gap-T}
	\text{type-I:} \ \Delta \sim \sqrt{T}; \ \text{type-II:} \ \Delta \sim T.
\end{align}
Eqs.~\eqref{eq:PG-gap-S} and \eqref{eq:PG-gap-T} are important in that they relate the pseudo-Goldstone gap to experimental signatures and in principle allow one to identify ObD in candidate materials. However, the Holstein-Primakoff approximation makes use of the large-$S$ limit, and it is not clear in what sense the results in Refs.~\cite{Rau2018,Khatua2023} depend on this limit or are universal.

In this work, we give an alternative formulation of ObD for systems with frustrated anisotropic spin interactions using spin functional integration. Apart from being conceptually simple, the spin functional integral does not require the large-$S$ limit, and includes quantum and thermal fluctuations on the same footing. It is found that the frustrated anisotropic spin terms responsible for ObD appear as gradient-dependent interactions of the pseudo-Goldstone modes, which vanish when the Lagrangian is minimized with a uniform spin ansatz. Accordingly the ground state sector acquires an `accidental' symmetry accompanied by gapless excitations. Fluctuations then generate a gradient-independent effective potential which removes the unphysical symmetry and gives a gap to the pseudo-Goldstone modes. 

We show that the leading contributions to the effective potential correspond to one-loop diagrams in the large-$S$ or the classical small temperature limits. The pseudo-Goldstone gap scalings in Eqs.~\eqref{eq:PG-gap-S} and \eqref{eq:PG-gap-T} are recovered respectively in these two limits. We demonstrate these conclusions by first explicitly computing $\Delta$ for the square compass (type-I) and cubic compass (type-II) models of ObD using second order perturbation theory. We then argue, based on scaling properties of type-I and type-II pseudo-Goldstone propagators, that Eqs.~\eqref{eq:PG-gap-S} and \eqref{eq:PG-gap-T} hold in the one-loop approximation for the effective potential, which is dominant in the two limits. Here the spin functional integral method offers a powerful yet simple framework that can take into account higher order corrections beyond the one-loop limit. Furthermore, it allows us to discuss other types of ObD by $m$-spin interactions where $m>2$. 

The rest of this paper is organized as follows. In Sec.~\ref{sec:spin-functional-integral} we review the functional integration formalism for spin systems. In Sec.~\ref{sec:cubic-compass} we study the cubic compass model and use perturbation theory to show that the pseudo-Goldstone gap is of type-II. We consider the type-I square compass model in Sec.~\ref{sec:square-compass}, and find agreements with the results in Ref.~\cite{Khatua2023} using the Holstein-Primakoff approximation. In Sec.~\ref{sec:ObD-scaling} we present general arguments for the pseudo-Goldstone gap dependence in Eqs.~\eqref{eq:PG-gap-S} and \eqref{eq:PG-gap-T} in the two aforementioned limits. We also predict other types of ObD due to general $m$-spin interactions. Finally in Sec.~\ref{sec:conclusion} we conclude by discussing possible research directions.

\section{Functional integration for spins}
\label{sec:spin-functional-integral}

First we review the functional integration formalism for spin systems which we shall use for subsequent analysis~\cite{Tsvelik2005,Auerbach2012}.

Spin-$S$ operators can be represented in terms of the two-component Schwinger bosons with spins up and down $b_{\pm}$:
\begin{equation}
	\bm{S}(\mathbf{r}) = \frac{1}{2} \sum_{\alpha,\beta = \pm}b_{\alpha}^\dagger (\mathbf{r}) \bm{\sigma}_{\alpha\beta} b_\beta(\mathbf{r}); \ \sum_{\alpha =\pm} b_{\alpha}^\dagger b_\alpha =2S.
	\label{eq::SchwingerBosons}
\end{equation}
The partition function of a given spin-$S$ system in $d$-dimension is:
\begin{equation}
	Z = \int D b b^\dagger \delta\left(\sum_{i} b_{i}^\dagger b_i -2S \right)\exp\left(-\int \mathrm{d}^dx \mathrm{d}\tau \mathcal{L}\right),
\end{equation}
where the Lagrangian:
\begin{equation}
	\mathcal{L} = b^\dagger \frac{\p}{\p \tau} b + H[b,b^\dagger].\label{eq::Lagrangian}
\end{equation}
$H$ is the spin Hamiltonian written in terms of the bosons $b_\pm$.

It is convenient for our purpose to choose the following parameterisation of $b_\pm$:
\begin{equation}
	b_+ = \sqrt{2S} \cos{\frac{\theta}{2}} e^{i(\chi-\phi)/2}, \ b_- = \sqrt{2S} \sin{\frac{\theta}{2}} e^{i(\chi+\phi)/2},
	\label{eq::EulerAngle}
\end{equation}
where $\theta,\pi/2-\phi, \chi$ form the Euler angles. The constraint in Eq.~(\ref{eq::SchwingerBosons}) is then satisfied identically, and the spin operator in Eq.~(\ref{eq::SchwingerBosons}) becomes:
\begin{equation}
	\bm{S} = S \mathbf{n} = S (\sin \theta \cos \phi,\sin \theta \sin \phi, \cos \theta).
\end{equation}
$\theta,\phi$ are the polar and azimuthal angles of the directional vector of the spin $\mathbf{n}$ in spherical coordinates. Substituting Eq.~(\ref{eq::EulerAngle}) into Eq.~(\ref{eq::Lagrangian}), the kinetic term is, up to a total derivative:
\begin{equation}\label{eq:kinetic-1}
	b^\dagger \frac{\p}{\p \tau} b = -i S \cos \theta \frac{\p \phi}{\p \tau}.
\end{equation}
The $\chi$ field drops out from both $\mathbf{n}$ and the kinetic term. This is because $\chi$ parameterizes rotations around the axis perpendicular to $\mathbf{n}$, and is thus unphysical. We then neglect $\chi$ below. Eq.~\eqref{eq:kinetic-1} suggests that the canonical momentum for $\phi$ is $S^z$. This is necessary for satisfying the spin commutation relation:
\begin{equation}
	[S^z,S^\pm] = \pm S^\pm.
\end{equation}


Eq.~\eqref{eq:kinetic-1} is usually inconvenient for practical purposes as it is written explicitly in $\phi,\theta$. To write it in vector form one introduces a new variable $u$ such that:
\begin{equation}
	\mathbf{n}(u=1,\tau)=\mathbf{n}(\tau), \ \mathbf{n}(0,\tau)=\text{const.}
\end{equation}
We then have:
\begin{equation}
	\begin{split}
		\int \mathrm{d}\tau \cos \theta \frac{\p \phi}{\p \tau}
		=& \int \mathrm{d}\tau \mathrm{d}u \left[ \frac{\p}{\p u} \left(\cos \theta \frac{\p \phi}{\p \tau}\right)- \frac{\p}{\p \tau} \left(\cos \theta \frac{\p \phi}{\p u}\right)  \right] \\
		=&-\int \mathrm{d}\tau \mathrm{d}u \sin \theta \left( \frac{\p\theta}{\p u}  \frac{\p \phi}{\p \tau}- \frac{\p\theta}{\p \tau}\frac{\p \phi}{\p u}  \right).
	\end{split}
\end{equation}
This is the 2D winding number $k$ on the manifold $(u,\tau)$. Therefore the kinetic term can be written in vector notation as:
\begin{equation}\label{eq::Kineticterm}
	W[\mathbf{n}] =  i S \int \mathbf{n}.\left( \frac{\p \mathbf{n}}{\p u} \times  \frac{\p \mathbf{n}}{\p \tau} \right) \mathrm{d}\tau \mathrm{d}u = 4\pi i S k.
\end{equation}
For both integer and half-integer $S$, $W[\mathbf{n}]$ is an integer multiple of $2\pi$.

The kinetic term as written in Eq.~\eqref{eq::Kineticterm} contains the additional unphysical variable $u$. The extra dimension drops out in calculations due to the following identity:
\begin{equation}\label{eq::Kineticterm-identity}
	\delta W[\mathbf{n}] =  i S \int \mathrm{d}\tau \delta\mathbf{n}.\left(\p_\tau \mathbf{n}  \times \mathbf{n}\right).
\end{equation}
This can be shown by substituting $\mathbf{n} \rightarrow \mathbf{n}+ \delta \mathbf{n}$ into the kinetic term in \eqref{eq::Kineticterm}.
\begin{equation}
	\delta \left[ \mathbf{n}.\left(\p_u \mathbf{n}  \times \p_\tau \mathbf{n} \right)\right]= \delta\mathbf{n}.\left(\p_u \mathbf{n}  \times \p_\tau \mathbf{n}\right)+\mathbf{n}.\left(\p_u \delta\mathbf{n}  \times \p_\tau \mathbf{n}\right)+\mathbf{n}.\left(\p_u \mathbf{n}  \times \p_\tau \delta\mathbf{n}\right).
\end{equation}
The first term vanishes since all three vectors lie inside the same plane due to the constraint $\mathbf{n}. \delta \mathbf{n}=0$. Integrating by parts in the two other terms give:
\begin{equation}\label{eq::Kineticterm-identity-1}
	\delta W[\mathbf{n}] =  i S \int \mathrm{d}\tau \mathrm{d}u \frac{\p}{\p u} \left[ \delta\mathbf{n}.\left(\p_\tau \mathbf{n} \times \mathbf{n}\right) \right] = i S \int \mathrm{d}\tau \delta\mathbf{n}.\left(\p_\tau \mathbf{n}  \times \mathbf{n}\right).
\end{equation}
For a ferromagnetic Hamiltonian with $\mathbf{n}$ slow varying in space, using Eq.~\eqref{eq::Kineticterm-identity-1} one gets the equation of motion:
\begin{equation}
	i S \left(\p_\tau \mathbf{n}  \times \mathbf{n} \right) = \frac{\delta H}{\delta \mathbf{n}} + \lambda \mathbf{n},
\end{equation}
where $\lambda$ is the Lagrange multipler for the constraint $\mathbf{n}^2=1$. Taking the vector product with $\mathbf{n}$ on the left gives the Landau-Lifshitz equation:
\begin{equation}
	i S \p_\tau \mathbf{n}  =\mathbf{n}\times \frac{\delta H}{\delta \mathbf{n}}.
\end{equation}

\section{Cubic Compass model}
\label{sec:cubic-compass}

\begin{figure}
	\centering
	\includegraphics[width=\linewidth]{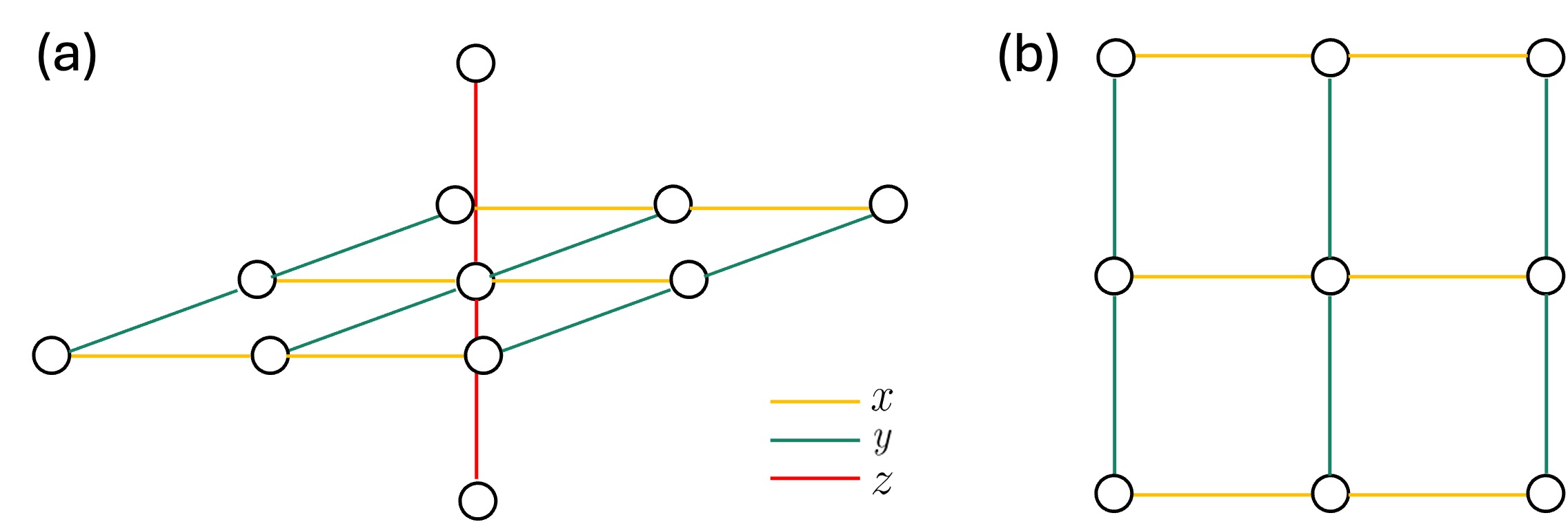}
	\caption{(a) the cubic lattice for the cubic compass model \eqref{eq:Hamiltonian-cubic-compass}. (b) the square lattice for the square compass model \eqref{eq:Hamiltonian-square-compass}. The nearest neighbour $x,y,z$ bonds are shown in orange, green and red respectively.}
	\label{fig:lattice}
\end{figure}

We start from the type-II ferromagnetic cubic compass model defined by the following Hamiltonian on a cubic lattice:
\begin{equation}
	H = - J \sum_{i,\alpha} \bm{S}(\mathbf{r}_i).\bm{S}(\mathbf{r}_i+\mathbf{e}_\alpha)  - K \sum_{i,\alpha} S^\alpha(\mathbf{r}_i)S^\alpha(\mathbf{r}_i+\mathbf{e}_\alpha).\label{eq:Hamiltonian-cubic-compass}
\end{equation}
where $\bm{S}(\mathbf{r}_i)$ are spin-$S$ operators at site $i$ and the index $\alpha=x,y,z$ sums over the nearest neighbour (N.N.) vectors with the corresponding FM Heisenberg $J>0$ and anisotropic interactions $K>0$;~see Fig.~\ref{fig:lattice}(a). The Hamiltonian is symmetric under combined spin and spatial $C_4$ rotations around $x$-, $y$-, $z$-axes, but the classical ground state is degenerate with respect to spin directions. Below we shall show that the 'accidental' spin rotation symmetry is removed by fluctuation corrections.

To study the fluctuation effects, we pass to a continuum Lagrangian density given by the local directional vector of the spin $\bm{S} = S \mathbf{n}(\mathbf{r})$. The N.N. terms are transformed as follows:
\begin{equation}
	S^\beta(\mathbf{r}) S^\beta(\mathbf{r}+\mathbf{e}_\alpha) = S^2\left\{[n^\beta(\mathbf{r})]^2 + n^\beta(\mathbf{r}) \p_i n^\beta(\mathbf{r})(\mathbf{e}_\alpha)_i a + \frac{1}{2} n^\beta(\mathbf{r}) \p^2_i n^\beta(\mathbf{r})a^2 \right\}.
\end{equation}
The linear term vanishes in summing over $\alpha$. The last term is transformed by integrating by parts and using:
\begin{equation}
	\sum_i f(\mathbf{r}_i) = \frac{1}{a^d} \int \mathrm{d}^dx f(x),
\end{equation}
where $a$ is the lattice constant. Unless stated otherwise, we shall take $a=1$. This gives the Lagrangian:
\begin{equation}
	\mathcal{L} = W[\mathbf{n}] + \frac{JS^2}{2}(\p_\mu \mathbf{n})^2 + \frac{KS^2}{2} \sum_\alpha\left(\frac{\p n^\alpha}{\p x^\alpha} \right)^2, \label{eq:Lagrangian-cubic-compass}
\end{equation}
where $W[\mathbf{n}]$ is the kinetic term given in \eqref{eq::Kineticterm}. The anisotropic N.N. terms are quadratic in gradient of $\mathbf{n}$, which vanishes when assuming $\mathbf{n}$ to be uniform. In what follows it will be shown that fluctuation corrections generate a gradient-independent effective potential which makes the pseudo-Goldstone modes massive. 

For this purpose, it is convenient to parameterize $\mathbf{n}$ as:
\begin{equation}
	\mathbf{n} = \left(\sqrt{1-\rho^2} \cos \phi,\sqrt{1-\rho^2} \sin \phi,\rho\right), \label{eq:param-cubic-compass}
\end{equation}
where $\phi$ is the in-plane polar angle of $\mathbf{n}$ and $\rho$ characterizes the out-of-plane oscillations. The integral measure in the functional integral is trivial:
\begin{equation}
	\int \mathrm{D}\mathbf{n} \delta(n^2-1) = \int \frac{\mathrm{D}n_x \mathrm{D}n_y}{\sqrt{1-n_x^2-n_y^2}} = \int \mathrm{D}\rho \mathrm{D}\phi,
\end{equation}
which simplifies the calculations. 

We first transform the kinetic term \eqref{eq::Kineticterm} under Eq.~\eqref{eq:param-cubic-compass}. Regarding $\rho$ as small oscillations, which will be justified below, we represent $\mathbf{n}$ to leading order in $\rho$ as:
\begin{equation}
	\mathbf{n} = \mathbf{n}_0 +\delta \mathbf{n}, \ \mathbf{n}_0 = (\cos \phi, \sin \phi ,0), \ \delta\mathbf{n}= (0,0, \rho),
\end{equation}
and substitute into the identity Eq.~\eqref{eq::Kineticterm-identity} to give:
\begin{equation}
	W = i S \int \mathrm{d}\tau \delta\mathbf{n}. (\p_\tau \mathbf{n}_0 \times\mathbf{n}_0 ) = i  S \rho \p_\tau \phi. 
	\label{eq:kinetic-cubic-compass}
\end{equation}
Substituting Eqs.~\eqref{eq:kinetic-cubic-compass} and 
\eqref{eq:param-cubic-compass} into the Lagrangian \eqref{eq:Lagrangian-cubic-compass} gives:
\begin{equation}
	\mathcal{L} =  i S  \rho \p_\tau \phi +\frac{JS^2}{2} \left[(\p_\mu \rho)^2 + (\p_\mu \phi)^2 \right]+\frac{JS^2}{2} \rho^2 \left[(\p_\mu \rho)^2 - (\p_\mu \phi)^2 \right]+V.\label{eq:Lagrangian-cubic-compass-1}
\end{equation}
The two terms in the middle are from the isotropic Heisenberg terms which do not generate a mass for the pseodu-Goldstone modes. The gradient-dependent anisotropic perturbation $V$ is: 
\begin{equation}
	\begin{split}
		V=& \frac{KS^2}{2} \Large\{(1-\rho^2)[ \sin^2 \phi (\p_x\phi)^2 +\cos^2 \phi (\p_y\phi)^2] + \rho  \cos \phi \sin \phi(\p_x\rho \p_x \phi - \p_y \rho \p_y \phi ) \\
		&+ \rho^2[\cos^2\phi(\p_x\rho)^2+\sin^2\phi(\p_y \rho)^2] + (\p_z \rho)^2\Large\}.
	\end{split} \label{eq:perturbation-cubic-compass}
\end{equation}
In our model, $V$ breaks the spin rotation symmetry but vanishes assuming a uniform spin ansatz. Therefore the magnons are gapless in the classical limit. However once fluctuations are included, they should generate an effective potential which determines the discrete ground state spin configurations and gaps the magnons. We shall show below that to second order in $K$, the effective potential is given by:
\begin{equation}\label{eq:perturbation-cubic-compass-rho}
	\Gamma=\frac{1}{2}A \rho^2-B \cos 4\phi, \ A,B>0.
\end{equation}
The coefficients $A,B$ further satisfy the following relations in the large $S$ and classical small temperature $T$ limits:
\begin{equation}\label{eq:perturbation-cubic-compass-scaling}
	A,B \sim S \ (\text{large-S}); \ A,B \sim T \ (\text{classical small-}T)
\end{equation}
Thus, the averaged spin lies along $x$- or $y$-axis: $\phi_0 =\langle \phi \rangle = 0, \pi/2, \pi, 3\pi/2$. It can also lie along the $z$-direction due to rotation symmetry. The psuedo-goldstone gap $\Delta$ is then found to agree with \eqref{eq:PG-gap-S} and \eqref{eq:PG-gap-T}.

To compute the effective potentials, we first need the free boson propagators of the Lagrangian \eqref{eq:Lagrangian-cubic-compass-1} defined as:
\begin{equation}
	\begin{split}
		D_0(i\omega_n,\mathbf{k}) &= \int \mathrm{d}\tau e^{i\omega_n\tau}
		\begin{pmatrix}
			\langle \text{T}_\tau \rho(\tau)\rho(0) \rangle & \langle\text{T}_\tau \rho(\tau)\phi(0) \rangle \\
			\langle\text{T}_\tau \phi(\tau)\rho(0) \rangle  & \langle\text{T}_\tau \phi(\tau)\phi(0) \rangle
		\end{pmatrix} \\
		&=
		\frac{1}{\omega_n^2 S^2 + J^2 S^4k^4}\begin{pmatrix}
			JS^2k^2 &  S \omega_n \\
			-S \omega_n & JS^2k^2
		\end{pmatrix}.
	\end{split}
\end{equation}
The bosonic Matsubara frequency $\omega_n = 2\pi n T, n \in \mathbb{Z}$. The pole gives the ferromagnet dispersion $\omega = JSk^2$.

\begin{figure}
	\centering
	\includegraphics[width=0.8\linewidth]{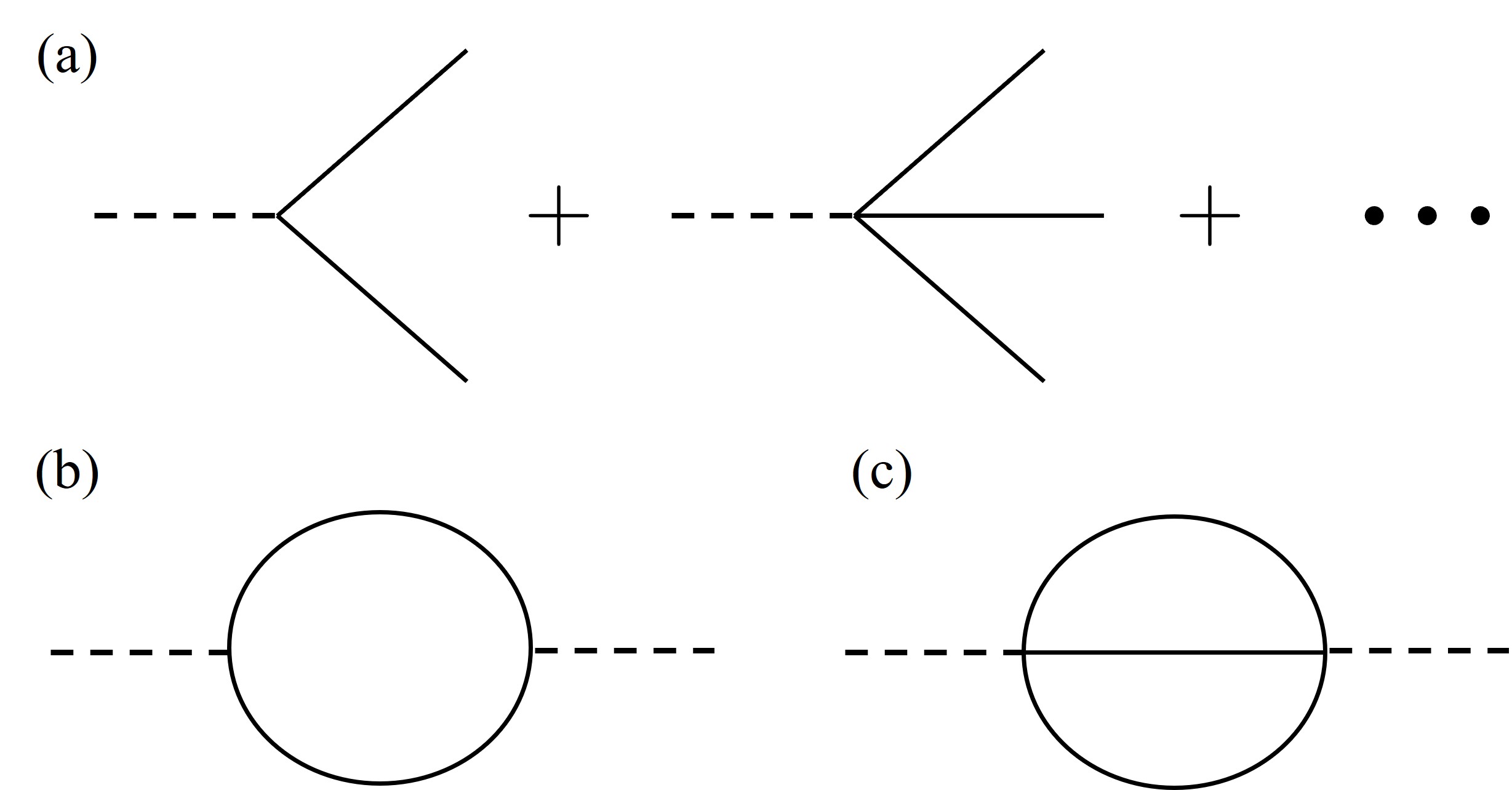}
	\caption{
		Diagrammatic representation of the cubic compass model Eq.~\eqref{eq:Lagrangian-cubic-compass-1}. (a) the interaction vertex corresponding to Eq.~\eqref{eq:vertex-expansion} for $n=0, 1$. The dashed lines represent the gradient-independent term given by the slow modes $\bm{\psi}_+$. (b) The second order effective potential where only the $n=0$ vertices enter. (c) The second order effective potential with $n=1$ vertices.}
	\label{fig:feynman-diagrams}
\end{figure} 

The rest of the calculations proceed in the standard way by separating the fast and slow modes for $\phi$ and $\rho$:
\begin{equation}\label{eq:fastmodes}
	\phi=\phi_-+\phi_+, \ \rho = \rho_-+ \rho_+,
\end{equation}
and integrate out the fast modes $\rho_+,\phi_+$. We substitute \eqref{eq:fastmodes} into \eqref{eq:perturbation-cubic-compass}. Since the interaction from the Heisenberg term is isotropic and does not generate a gap for the psuedo-Goldstone modes, they are neglected in our calculations. Since we wish to obtain a gradient-independent potential, in the interaction vertex Eq.~\eqref{eq:perturbation-cubic-compass} the fast modes should enter both derivatives. In the rest we expand in a Taylor series of fast modes. As can be seen from Eq.~\eqref{eq:perturbation-cubic-compass}, the relevant interaction vertex can be written in the general form:
\begin{equation}\label{eq:vertex-expansion}
	\sum_n \frac{1}{n!}\frac{\mathrm{d}^nf_{ij}^{ab}(\bm{\psi}_-) }{\mathrm{d}\psi^{i_1}_-...\mathrm{d}\psi^{i_n}_-}\left(\psi^{i_1}_{+}...\psi^{i_n}_+ \right)\p_i\psi^a_+ \p_j\psi^b_+,
\end{equation}
where $\bm{\psi}=(\rho,\phi)$ and $f_{ij}^{ab}(\bm{\psi})$ are the parts of Eq.~\eqref{eq:perturbation-cubic-compass} that do not contain derivatives. The $n=0$ vertex is shown in Fig.~\ref{fig:feynman-diagrams}(a), where the two solid lines correspond to spatial derivatives of $\psi_+$ and $f(\bm{\psi}_-)$ enters as an external potential with dashed lines. The $n=1$ vertex contains three fast modes and is shown in Fig.~\ref{fig:feynman-diagrams}(a) with three solid lines etc. In expanding the perturbation series of $V$ in powers of $\rho_+, \phi_+$ and integrating them out, the corresponding solid lines form closed loops, i.e. are contracted. Below we shall use the term contraction in this sense.

Generally Eq.~\eqref{eq:vertex-expansion} gives an infinite number of vertices with at least two solid lines corresponding to the gradient terms. Here we find two limits where the diagrammatic series is controlled: the large-$S$ limit and the classical limit at low temperature $T$. Both limits correspond to the one-loop expansion as will be discussed below.

Let us first discuss the large-$S$ limit. Consider e.g. the second order diagram given by $n=0$ in Fig.~\ref{fig:feynman-diagrams}(b), where only the gradients of $\psi_+$ are contracted, and also the diagram in Fig.~\ref{fig:feynman-diagrams}(c) for $n=1$ in which an additional contraction of $\psi_+$ fields occurs. In the large-$S$ limit, we have for the first diagram (neglecting frequency summation):
\begin{equation}
	I_0 \sim S^4 T \sum_n\frac{J^2S^4k^4k_y^2}{(\omega_n^2S^2 + J^2S^4k^4 )^2} \sim T \sum_n\frac{J^2k^4k_y^2}{[(\omega_n/S)^2 + J^2k^4 ]^2} \sim S \int\mathrm{d}\omega\frac{ J^2k^4k_y^2}{(\omega^2+ J^2k^4)^2}.
\end{equation}
Here we have used:
\begin{equation}
	\lim_{\varepsilon\rightarrow0} \sum_n f(n \varepsilon) = \frac{1}{\varepsilon}\int f(x)\mathrm{d}x.  \label{eq:integral-def}
\end{equation}
Thus each propagator contains a power of $S^{-2}$ whereas the summation over Matsubara frequency gives another factor of $S$. For the second diagram, we have one more power of the $\phi$ propagator, and another internal Matsubara frequency summation. The former gives $S^{-2}$ factor while the latter gives another factor of $S$:
\begin{equation}
	I_1 \sim O(S).
\end{equation}
Thus the second diagram is sub-leading as $S\rightarrow\infty$. In Sec.~\ref{sec:ObD-scaling} we show generally that diagrams given by $n>2$ vertices are subleading.

Similarly in the classical limit, the summation over Matsubara frequency becomes:
\begin{equation}\label{eq:Matsubara-sum-classical}
	T \sum_n f(\omega_n) \rightarrow Tf(0).
\end{equation}
Thus at small temperature $T$, the diagrammatic series is controlled by the number of Matsubara summations. Again we find that $n=0$ vertices are dominant with 
\begin{equation}
	I_0 \sim T,
\end{equation}
and $n>0$ vertices are subleading as $T\rightarrow 0$.

\begin{figure}
	\centering
	\includegraphics[width=\linewidth]{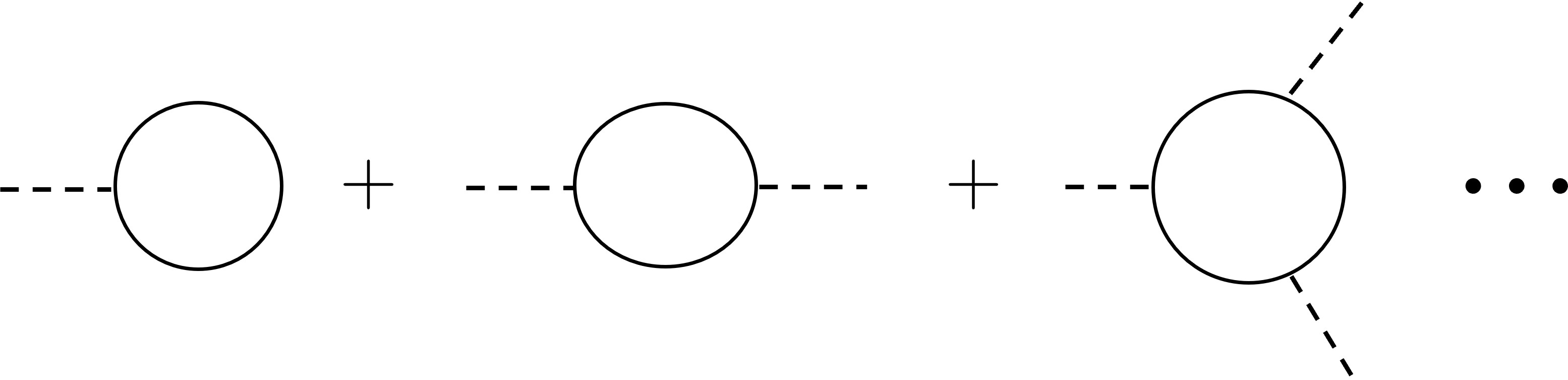}
	\caption{
		One-loop expansion of the gradient-dependent vertices Eq.~\eqref{eq:vertex-general}.}
	\label{fig:one-loop}
\end{figure}

We now compute the effective potential in the two aforementioned limits. In the cubic compass model, $K$ is small and perturbation theory can be applied. Thus in what follows we shall only find the effective potential up to order $K^2$ . In Sec.~\ref{sec:ObD-scaling} we shall show that, even if $K$ is not small, the leading contributions in the two limits are the one-loop diagrams as shown in Fig.~\ref{fig:one-loop}, and the results for the pseudo-Goldstone gap are general in the two limits. 

First we show that the $\rho$ mode acquires a mass by rewriting the second line in Eq.~\eqref{eq:perturbation-cubic-compass}:
\begin{equation}
	\begin{split}
		\frac{KS^2}{4} \left\{\rho_-^2[(1+\cos 2\phi_-)(\p_x\rho_+)^2+(1-\cos 2\phi_-)(\p_y \rho_+)^2] + (\p_z \rho_+)^2\right\}.
	\end{split}
\end{equation}
This gives to leading order:
\begin{equation}\label{eq:rho-gap}
	\frac{1}{2}\Gamma_\rho \rho_-^2 = \frac{KS^2}{4}\rho_-^2 \langle(\p_x\rho_+)^2 + (\p_y\rho_+)^2 \rangle = \frac{KS^4}{4}\rho_-^2 T\sum_n \sum_{\mathbf{k}}\frac{Jk^2(k_x^2+k_y^2)}{S^2 \omega_n^2+J^2k^4S^4},
\end{equation}
which is described by the first diagram in Fig.~\ref{fig:one-loop} and after summing over $\omega_n$ has the form Eqs.~\eqref{eq:perturbation-cubic-compass-rho} and \eqref{eq:perturbation-cubic-compass-scaling}. Note that $\rho$ only acquires a mass for $K>0$. For the antiferromagnetic cubic compass model $K<0$, the $\rho$ mode condenses corresponding to the averaged spin lying out of plane, and the parameterization \eqref{eq:param-cubic-compass} becomes inadequate. 

We now compute the effective potential for $\phi$. Since $K\ll J$, we can use perturbation theory for the gradient-dependent interaction $V$ in Eq.~\eqref{eq:perturbation-cubic-compass}. Here up to $K^2$ in the one-loop approximation, the relevant term in the effective potential for $\phi$ is: 
\begin{equation}\label{eq:perturbation-cubic-compass-1}
	\frac{KS^2}{2} [ \sin^2 \phi_- (\p_x\phi_+)^2 +\cos^2 \phi_- (\p_y\phi_+)^2].
\end{equation}
The terms proportional to $\rho \p_i \phi \p_j \rho$ in Eq.~\eqref{eq:perturbation-cubic-compass} involve three propagators and are subleading. This gives the effective potential for the $\phi$-mode corresponding to Fig.~\ref{fig:feynman-diagrams}(b):
\begin{equation}
	\begin{split}
		\Gamma_\phi = - \frac{K^2S^4}{8} \bigg\{\sin^4 \phi_- Q_{xx} + \cos^4 \phi_- Q_{yy}   +2  \cos^2 \phi_-  \sin^2 \phi_- Q_{xy}\bigg\},
	\end{split}
\end{equation}
where we have defined the tensor:
\begin{equation}
	Q_{ab} = \int \mathrm{d}^3x\left[\langle \p_a \phi_+(\mathbf{r})\p_b \phi_+(0) \rangle\right]^2, \ Q_{xx}=Q_{yy}= 3Q_{xy}= Q>0.
\end{equation}
The latter identity follows from the in-plane isotropy of the free Lagrangian. Thus the effective potential for the $\phi$-mode is:
\begin{equation}\label{eq:perturbation-cubic-compass-2}
	\Gamma_\phi =  -\frac{K^2S^4Q}{48} \cos 4\phi.
\end{equation}
From the previous discussions on second order diagrams, $Q$ scales as $S$ or $T$ in the large-$S$ and classical small-$T$ limit respectively. As a result $\Gamma_\phi$ has the form given by Eqs.~\eqref{eq:perturbation-cubic-compass-rho} and \eqref{eq:perturbation-cubic-compass-scaling}. The psuedo-goldstone gap $\Delta$ is found by expanding Eq.~\eqref{eq:perturbation-cubic-compass-2} to quadratic order around $\phi_0$ and inverting the renormalized propagator at zero momentum after replacing $i\omega_n \rightarrow \Delta$:
\begin{equation}
	\left[D(i\omega=\Delta,0)\right]^{-1}=
	\begin{pmatrix}
		\Gamma_\rho'' & -iS\Delta \\
		i S \Delta  & \Gamma_{\phi_0}^{''}
	\end{pmatrix} =0 ,\  \Gamma_{\rho}^{''}= \frac{\p^2 \Gamma_\rho}{\p \rho^2},\  \Gamma_{\phi_0}^{''}= \frac{\p^2 \Gamma_\phi}{\p \phi_0^2}.
\end{equation}
Substituting \eqref{eq:perturbation-cubic-compass-scaling} we recover the  type-II pseudo-Goldstone gap in Eqs.~\eqref{eq:PG-gap-S} and \eqref{eq:PG-gap-T}:
\begin{equation}\label{eq:PG-gap-II}
	\Delta \sim O(S) \ (\text{large-}S), \  \Delta \sim T \ (\text{classical low-}T).
\end{equation}

\section{Square Compass model}
\label{sec:square-compass}

We now consider another example belonging to type-I ObD: the ferromagnetic square compass model~\cite{Rau2018,Khatua2023}. As shown in Eq.~\eqref{eq:dispersion-square-compass} below, even though the system is ferromagnetic, the single pseudo-Goldstone mode has linear dispersion. The Hamiltonian is given by \eqref{eq:Hamiltonian-cubic-compass} but on a 2D square lattice:
\begin{equation}\label{eq:Hamiltonian-square-compass}
	H = - J \sum_{i,\alpha} \bm{S}(\mathbf{r}_i).\bm{S}(\mathbf{r}_i+\mathbf{e}_\alpha)  - K \sum_{i} \left[ S^x(\mathbf{r}_i)S^x(\mathbf{r}_i+\mathbf{e}_x)+ S^y(\mathbf{r}_i)S^y(\mathbf{r}_i+\mathbf{e}_y) \right].
\end{equation}
The summation is taken over N.N. vectors with index $\alpha = x, y$, and the exchange integrals satisfy $K\ll J$, as shown in Fig.~\ref{fig:lattice}(b). Here the system symmetry is combined spin and spatial $C_4$ rotations along $x$-, $y$-axes. As in Sec.~\ref{sec:cubic-compass}, we pass to the continuum Lagrangian:
\begin{equation}
	\mathcal{L} = W[\mathbf{n}] + \frac{JS^2}{2} (\p_\mu \mathbf{n})^2 + KS^2(n^z)^2 +V, \ V=\frac{KS^2}{2} \sum_{\alpha=x,y}\left(\frac{\p n^\alpha}{\p x^\alpha} \right)^2. \label{eq:Lagrangian-square-compass}
\end{equation}
In addition to the gradient-dependent interactions, the system has an easy-plane anisotropy on the classical level which gaps out the out-of-plane oscillations..

We shall now derive the pseudo-Goldstone gap for the square-compass-model by regarding $K\ll J$ as a small parameter. Let us first discuss the renormalization effect in two-dimensions which is significant. For example consider a Heisenberg ferromagnet at finite temperatures. Under the renormalization group (RG), the exchange integral $J$ is renormalized to zero at a finite energy scale, due to the strong interaction of Goldstone modes that restore the $O(3)$ spin rotation symmetry in the infra-red limit~\cite{Polyakov1975}. However, here it is expected that the pseudo-Goldstone gap generated by the fluctuations will regularize the infra-red divergences and the system should retain magnetic order, as the system has discrete symmetry only. Therefore, in what follows we shall not consider the RG effect.

Using again the parameterization \eqref{eq:param-cubic-compass}, the continuum Lagrangian becomes:
\begin{equation}
	\mathcal{L} =  i S  \rho \p_\tau \phi +\frac{JS^2}{2} \left[(\p_\mu \rho)^2 + (\p_\mu \phi)^2 \right]+ K S^2\rho^2 +\frac{JS^2}{2} \rho^2 \left[(\p_\mu \rho)^2 - (\p_\mu \phi)^2 \right]+V,\label{eq:Lagrangian-square-compass-1}
\end{equation}
where the gradient-dependent interaction is again given by \eqref{eq:perturbation-cubic-compass} without the $(\p_z \rho)^2$ term.

The free boson propagator is:
\begin{equation}
	D(i\omega_n,\mathbf{k}) = 
	\frac{1}{\omega_n^2 S^2 + J S^4k^2(Jk^2 + 2K)}\begin{pmatrix}
		JS^2k^2 &  S \omega_n \\
		-S \omega_n & JS^2k^2+ 2KS^2
	\end{pmatrix}.
\end{equation}
The dispersion is linear at small momentum:
\begin{equation}
	\omega = S\sqrt{J (Jk^2 + 2K)}k.\label{eq:dispersion-square-compass}
\end{equation}

The rest of the calculations are similar to those in Sec.~\ref{sec:cubic-compass} (the dimensionality of the system does not enter into the analysis there). The dominant perturbation is again given by Eq.~\eqref{eq:perturbation-cubic-compass-1}, which gives the effective potential for $\phi$ Eq.~\eqref{eq:perturbation-cubic-compass-2}. Here the difference from the type-II cubic compass model is that $\Gamma_\rho\sim S^2$ instead of being linear in $S$, since this term is not generated by the fluctuations. Thus the averaged spin lies in-plane along the $x$-, $y$-directions. Repeating the same arguments as in Sec.~\ref{sec:cubic-compass}, the pseudo-Goldstone gap is found from the pole of the boson propagator at zero momentum:
\begin{equation}
	\left[D(i\omega=\Delta,0)\right]^{-1}=
	\begin{pmatrix}
		\Gamma_\rho'' & -iS\Delta \\
		i S \Delta  & \Gamma_{\phi_0}^{''}
	\end{pmatrix} =0 ,\ \Gamma_\rho'' = 2KS^2,\ \Gamma_{\phi_0}^{''}= \frac{\p^2 \Gamma_\phi}{\p \phi_0^2}\sim S. 
\end{equation}
We then recover exactly the type-I pseudo-Goldstone gap in Eqs.~\eqref{eq:PG-gap-S} and \eqref{eq:PG-gap-T}:
\begin{equation}\label{eq:PG-gap-I}
	\Delta \sim  \sqrt{S} \ (\text{large-}S), \  \Delta \sim  \sqrt{T} \ (\text{classical low-}T).
\end{equation}
Finally we note that the temperature scaling and averaged spin directions also agree with the classical Monte Carlo simulation in Ref.~\cite{Khatua2023} at small temperatures. 

It will be shown in Sec.~\ref{sec:ObD-scaling} that Eq.~\eqref{eq:PG-gap-I} is due to the linear dispersion of the free boson propagator Eq.~\eqref{eq:dispersion-square-compass} at small momenta. For a Heisenberg anti-ferromagnet, the spin-wave also has linear dispersion. In fact, we shall show in Appendix~\ref{sec:easyplane} that in the easy-plane limit $K \gg J$, the Lagrangian~\eqref{eq:Lagrangian-square-compass} has the same form as a planar AFM Lagrangian, and the pseudo-Goldstone gap relation Eq.~\eqref{eq:PG-gap-II} still holds. Thus we support the statement that ObD of AFM is generally of type-I~\cite{Rau2018}.

\section{Scaling arguments for ObD}
\label{sec:ObD-scaling}

We have demonstrated the pseudo-goldstone gap relations Eqs.~\eqref{eq:PG-gap-S} and \eqref{eq:PG-gap-T} in Secs.~\ref{sec:cubic-compass} and \ref{sec:square-compass} by explicit examples. We now prove that Eqs.~\eqref{eq:PG-gap-S} and \eqref{eq:PG-gap-T} hold more generally using scaling arguments of the free massless pseudo-Goldstone propagators.

As we have seen in Secs.~\ref{sec:cubic-compass} and \ref{sec:square-compass}, the spin-wave oscillations can be characterized by a collection of real fields $\bm{\psi}$, e.g. $\phi$ and $\rho$ in the previous sections for collinear ferromagnets. The anisotropic perturbations in ObD are generally an expansion in gradients of $\bm{\psi}$. Since linear gradient terms imply an instability for the slow-varying spin configurations, the anisotropic perturbations must start from quadratic gradient terms of the form [c.f. Eq.~\eqref{eq:vertex-expansion}]:
\begin{equation}\label{eq:vertex-general}
	V=f_{ij}^{ab}(\bm{\psi}) \p_a \psi^i \p_b \psi^j,
\end{equation}
where $f_{ij}^{ab}$ are certain gradient-independent functions of $\bm{\psi}$ and are proportional to $ S^2$ due to that two-spin interactions in the Hamiltonian. Eq.~\eqref{eq:vertex-general} is represented by Fig.~\ref{fig:feynman-diagrams}(a) with two solid lines corresponding to gradient terms.

Generally, the anisotropic vertex~\eqref{eq:vertex-general} generates a gradient-independent effective potential due to fast mode fluctuations. But Type-I and type-II ObD differ in the form of their free massless propagators of the pseudo-Goldstone modes. Note that this excludes the $\rho$-mode in the square compass model Sec.~\ref{sec:square-compass} since it has a gap in the free Lagrangian; c.f. also Appendix~\ref{sec:easyplane} where the $\rho$-mode is integrated out in the limit $K\gg J$. Let us first discuss the structure of the propagators and their scaling in $S$. For a ferromagnet-like (type-II) Lagrangian, the kinetic part scales as $i\omega_n S$. The dispersion~\eqref{eq:ObD-dispersion} gives for the propagator:
\begin{equation}
	D_0(i\omega_n,\mathbf{k}) \sim (i\omega_n S - S^2k^2)^{-1} \sim S^{-2} f\left(\frac{\omega_n}{S} \right). \label{eq:propagator-II}
\end{equation}
For the type-I Lagrangian, the propagator corresponding to the dispersion Eq.~\eqref{eq:ObD-dispersion} is:
\begin{equation}
	D_0(i\omega_n,\mathbf{k}) \sim (\omega_n^2 + S^2k^2)^{-1} \sim S^{-2} f\left(\frac{\omega_n}{S} \right). \label{eq:propagator-I}
\end{equation}
Here we note that the summation over Matsubara frequency in the large-$S$ limit:
\begin{equation}\label{eq:summation-matsubara}
	\lim_{S\rightarrow \infty} T \sum_n g\left(\frac{\omega_n}{S}\right) = \frac{S}{2\pi} \int \mathrm{d}x g\left(x\right). 
\end{equation}
Therefore, for both type-I and type-II cases, each summation over Matsubara frequency brings a single power of $S$.

We now show that the argument given in Sec.~\ref{sec:cubic-compass} for restricting to the one-loop expansion in the large-$S$ limit is general. Consider the $n$-th order diagram in the one-loop expansion; see Fig.~\ref{fig:feynman-diagrams}(d) which shows the diagrams for $n\le3$. There are $n$ vertices which carry power $S^{2n}$, which is cancelled by $S^{-2n}$ due to $n$ propagators. Finally there is a factor $S$ from summing over the single loop Matsubara frequency. Beyond one-loop, i.e. we have $m$ additional integrations over internal frequencies arising from $m$ additional propagators. The former scales as $S^{m}$ whereas the latter gives $S^{-2m}$. Thus we see that the dominant diagrams as $S \rightarrow\infty$ correspond to the one-loop expansion with scaling $\Gamma \sim S$. Hence, only the external lines corresponding to the gradients in Eq.~\eqref{eq:vertex-general} are contracted.

Similarly, in the classical limit at small temperature $T$, each sum over Matsubara frequency gives the small factor $T$; see Eq.~\eqref{eq:Matsubara-sum-classical}. The diagrammatic expansion in powers of $T$ is equivalent to expansion in number of internal frequency summations. Accordingly, the one-loop expansion is leading which scales as $\Gamma \sim T$.

So far we have shown that for both type-I and type-II, the effective potential $\Gamma$ has the scaling:
\begin{equation}\label{eq:scaling}
	\Gamma \sim S \ (\text{large-}S), \  \Gamma \sim T \ (\text{classical low-}T).
\end{equation}
For the pseudo-Goldstone gap $\Delta$, we replace the frequency $i\omega\rightarrow \Delta$ in the boson propagators in Eqs.~\eqref{eq:propagator-II} and \eqref{eq:propagator-I}, and include the gap generated by the effective potential. $\Delta$ is then determined by the pole of the propagator:
\begin{equation}
	\text{type-I:} \ \Delta^2 \sim \Gamma, \ \text{type-II:} \ S \Delta \sim \Gamma,
\end{equation}
which immediately gives the pseudo-Goldstone gap relations Eqs.~\eqref{eq:PG-gap-S} and \eqref{eq:PG-gap-T}.


We also obtain an explicit expression for the effective potential in the one-loop approximation in terms of the magnon dispersions, which can be found from e.g. linear spin-wave theory. This will be convenient in numerical evaluations. This can be done by integrating out the gradient terms in Eq.~\eqref{eq:vertex-general} by regarding $f_{ij}^{ab}(\bm{\psi})$ as external potentials:
\begin{equation}
	\Gamma = \frac{T}{2} \sum_{a,n,\mathbf{k}}\log \left( \omega_n^2 + \varepsilon_{a\mathbf{k}}^2\right), \label{eq:effective-potential-one-loop}
\end{equation}
where $\varepsilon_{a\mathbf{k}}(\bm{\psi})$ is the $a$-th pseudo-Goldstone dispersion. The summation over the Matsubara frequency gives the standard free energy:
\begin{equation}\label{eq:free-energy-one-loop}
	F(\bm{\psi}) = T \sum_{a,\mathbf{k}} \log \sinh \frac{\varepsilon_{a}(\mathbf{k})}{2T}.
\end{equation}
At zero temperature Eq.~\eqref{eq:free-energy-one-loop} reduces to the zero-point energy:
\begin{equation}
	E(\bm{\psi}) =  \frac{1}{2}\sum_{a,\mathbf{k}}\varepsilon_{a}(\mathbf{k}).
\end{equation}
Expanding around the minimum $\bm{\psi}$ values to quadratic order, we recover the pseudo-goldstone gap expressed in terms of the curvature of the zero-point energy~\cite{Rau2018}.

Minimizing Eq.~\eqref{eq:free-energy-one-loop} in the classical limit \eqref{eq:Matsubara-sum-classical} $T\rightarrow \infty$ is equivalent to minimizing the free energy:
\begin{equation} 
	F = T \sum_{a,\mathbf{k}}\log \varepsilon_{a\mathbf{k}},
\end{equation}
which has been used to find the equilibrium spin configuration in thermal ObD~\cite{Khatua2023}.

Are there other types of ObD from the ones considered so far? From the above discussions, it is clear that the type-I and -II ObD differ in their kinetic terms $W[\mathbf{n}]$ given by Eq.~\eqref{eq::Kineticterm}. The fluctuations from the microscopic spin Hamiltonian give identical scaling \eqref{eq:scaling} in both cases. However, Eq.~\eqref{eq:scaling} is based on $2$-spin interactions. For an Hamiltonian with $m$-spin interactions only where $m>2$, it is expected that the vertex still starts with quadratic powers of gradients like \eqref{eq:vertex-general}. But instead of Eqs.~\eqref{eq:propagator-II} and \eqref{eq:propagator-I} we have for the pseudo-Goldstone propagators:
\begin{equation}
	\text{type-I:} \ D_0(i\omega_n,\mathbf{k}) \sim S^{-m} f\left(\frac{\omega_n}{S^{m/2}} \right)  ; \ \text{type-II:} \  D_0(i\omega_n,\mathbf{k}) \sim S^{-m} f\left(\frac{\omega_n}{S^{m-1}} \right).
\end{equation}
In the large-$S$ limit, the same arguments as following \eqref{eq:summation-matsubara} show that the effective potential is given by the one-loop expansion, the other diagrams being subleading in powers of $S^{-m/2}$ (type-I) or $S^{-1}$ (type-II). Summing over the loop Matsubara frequency using Eq.~\eqref{eq:integral-def}, the one-loop effective potential has scaling:
\begin{equation}
	\text{type-I:} \ \Gamma \sim S^{m/2}, \ \text{type-II:} \ \Gamma \sim S.
\end{equation}
which gives the pseudo-Goldstone gap:
\begin{equation}\label{eq:PG-new}
	\text{type-I:} \ \Delta \sim S^{m/4}, \ \text{type-II:} \ \Delta \sim O(S).
\end{equation}
Therefore the pseudo-Goldstone gap acquires a different scaling. The temperature dependence in the classical small $T$ limit is unaffected.

The situation could also arise that only the ObD vertices are due to $m$-spin interactions, whereas the free pseudo-Goldstone propagators retain the form of Eqs.~\eqref{eq:propagator-II} and \eqref{eq:propagator-I} due to $2$-spin terms. In this case, each vertex is proportional to $S^{m}$ whereas each propagator gives $D_0(k)\sim S^{-2}$. The one-loop expansion gives (we omit loop momentum integration):
\begin{equation}
	\Gamma = T \sum_{n} \sum_{p=1}^{\infty}\frac{(-C)^p}{p}\left[D_0(i\omega_n) S^{m}\right]^p = T \sum_{n}\log[1+CD_0(i\omega_n)S^m]\sim (m-2)S\log S; \ m> 2.
\end{equation}
Here $C$ is a constant. The linear $S$ factor is due to the summation over the Matsubara frequency $\omega_n$ in \eqref{eq:summation-matsubara}. The factor $1/p$ is the standard coefficient for the $p$-th order diagrammatic expansion of the free energy~\cite{abrikosov2012methods}. The pseudo-Goldstone gap in the large-$S$ limit is:
\begin{equation}\label{eq:PG-new-1}
	\text{type-I:} \ \Delta \sim \sqrt{S\log S}, \ \text{type-II:} \ \Delta \sim \log S.
\end{equation}
Note that the scaling of $\Delta$ does not depend on $m$ as long as $m>2$.

\section{Conclusion}
\label{sec:conclusion}

In this work, we studied ObD due to frustrated anisotropic $2$-spin interactions and the associated pseudo-Goldstone gap using the spin functional integration method. We showed that the free Lagrangian has an unphysical continuous symmetry and the bare pseudo-Goldstone mode is gapless. The continuous symmetry is removed by the gradient-dependent interactions which however vanish in the uniform spin approximation. Fluctuations of the pseudo-Goldstone modes then generate a gradient-independent effective potential which removes the `accidental' degeneracy and the pseudo-Goldstone modes acquire a gap. 

We compute the pseudo-Goldstone gap explicitly for the cubic and square compass models, which belong to type-II and type-I ObD classified in Ref.~\cite{Rau2018}. In particular, our results on the pseudo-Goldstone gap dependence on $S$ and $T$ using second-order perturbation theory agree with Ref.~\cite{Rau2018} in the large-$S$ and the classical low-$T$ limits respectively. We then show that in these two limits, the leading diagrams for the effective potential are given by the one-loop expansion, and type-I and type-II ObD correspond to linear and quadratic dispersions of the bare pseudo-Goldstone modes. Thus the ferromagnetic cubic compass model belongs to type-II, whereas type-I is given by the easy-plane ferromagnetic square compass model. We then verify the gap dependence on $S$ and $T$ without using perturbation theory.

The spin functional integral allows us to straightforwardly extend to models with $m$-spin interactions $m>2$. We predict different dependence of the pseudo-Goldsone gap on $S$ in the large-$S$ limit given by Eq.~\eqref{eq:PG-new} for an $m$-spin Hamiltonian, and by Eq.~\eqref{eq:PG-new-1} where only the anisotropic vertices are given by $m$-spin interactions.

Our work showed that the semi-classical formula for the pG gap based on non-linear spin wave theory given in Ref.~\cite{Rau2018} corresponds to the one-loop approximation, which requires the large-$S$ approximation. This perhaps explains the fact that, for many materials with ObD, the theoretically predicted pG gaps are of the same of order of magnitude as the experimentally measured values, but they do not agree quantitatively~\cite{Rau2018}.

Several questions remain open from our analysis. Experimentally, disorder is present in the system and its effect on the pG gap can be important. This can be studied using spin functional integration by introducing random pG interactions. On the theoretical front, it remains interesting to extend the spin functional integral to AFM ObD, which is shown to be type-I using the Holstein-Primakoff approximation~\cite{Rau2018}. Secondly we have only considered collinear magnets. The application of spin functional integral to non-collinear magnets, where the pseudo-Goldstone modes are also linear in momenta, remain to be explored. Lastly, the ObD considered in this paper arises from frustrated anisotropic spin-interactions due to spin-orbit coupling. ObD also occurs from exchange frustration with spin isotropic interactions with competing classical ground states, e.g. the $J_1-J_2$ model on the triangular and square lattice. There, the pG modes coexist with true Goldstone modes of the magnetic order. For example, in the square lattice $J_1-J_2$ model in the $J_1<2J_2$ stripe phase with an ordering wave vector at $(0,\pi)$, there are true Goldstone modes at $(0, \pi)$ and $(0,0)$ but a pG mode appears at $(\pi,0)$ corresponding to the other stripe pattern. This differs from our case in several aspects: first the spin rotation symmetry is not broken; secondly the pG modes are not near the ordering momenta so the gradient expansion is invalid; both true Goldstone and pG modes coexist in the spectrum. The clarification of these technical challenges will be an important direction for future research.

\section*{Acknowledgements}
P. R. would like to thank M. Ciarchi, J. Habel, L. Mangeolle, J. Rau and Z.J. Wang for discussions on various aspects of this work.

\paragraph{Funding information}
JK acknowledges support from the Deutsche Forschungsgemeinschaft (DFG, German Research Foundation) under Germany’s Excellence Strategy (EXC–2111–390814868 and ct.qmat EXC-2147-390858490), and DFG Grants No. KN1254/1-2, KN1254/2-1 TRR 360 - 492547816 and SFB 1143 (project-id 247310070), as well as the Munich Quantum Valley, which is supported by the Bavarian state government with funds from the Hightech Agenda Bayern Plus. J.K. further acknowledges support from the Imperial-TUM flagship partnership.

\appendix

\section{Effective Lagrangian for the square compass model}
\label{sec:easyplane}

In this Appendix, we shall consider the spin-wave spectrum at small momentum for the square compass model. It will be shown that the effective Lagrangian resembles that of a planar AFM with gradient-dependent anisotropic interactions and the pseudo-Goldstone mode is still given by Eq.~\eqref{eq:PG-gap-II}. To focus on the linear dispersion at small momentum, we shall take the easy-plane limit $K\gg J$, which might arise as a result of renormalisation. This allows us to integrate out-of-plane oscillations explicitly and obtain an effective Lagrangian in the pseudo-Goldstone mode $\phi$ only.

Returning to the Lagrangian \eqref{eq:Lagrangian-square-compass}, we shall again use the parameterization \eqref{eq:param-cubic-compass} and derive the effective easy-plane Lagrangian. Here the easy-plane limit cannot be taken by simply setting $\mathbf{n}$ to be in-plane because the kinetic term $W[\mathbf{n}]= iS \rho \p_\tau \phi $ vanishes. Thus the $\rho$ fluctuations must be properly taken into account. For generality we shall derive the effective Lagrangian for an arbitrary easy plane ferromagnet. The easy-plane limit means that the Hamiltonian $H$ has a strong minimum around $\rho =0$. Therefore in the partition function only small $\rho$ is important for the functional integral, and one can use the the saddle-point approximation:
\begin{equation}
	\begin{split}
		Z&\approx \int \mathrm{D}\rho \mathrm{D}\phi \exp\left[- \int \mathrm{d}\tau \mathrm{d}^2 x \left(iS \rho \p_\tau \phi + H_0+ \frac{1}{2}H^{''}_0 \rho^2]\right)\right]
		\\
		&\propto  \int \mathrm{D}\phi \exp\left[- \int \mathrm{d}\tau \mathrm{d}^2 x \left( \frac{S^2}{2|H^{''}_0 |} (\p_\tau\phi)^2 + H_0 \right)\right],
	\end{split}
\end{equation}
where $H_0$ is the Hamiltonian with $\mathbf{n}$ in-plane. We see that generally in the easy-plane limit, due to the transverse fluctuations the spin-wave dispersion becomes linear. In the square compass model $H^{''}_0 = 2KS^2$ to leading order, and we obtain the effective Lagrangian:
\begin{equation}
	\mathcal{L}_{\text{eff}} 
	=\frac{1}{4K} (\p_\tau\phi)^2 + \frac{JS^2}{2} (\p_\mu \phi)^2+ \frac{K^{'}S^2}{2}\left[\sin^2 \phi (\p_x\phi)^2+\cos^2 \phi (\p_y\phi)^2 \right].\label{eq:effective-Lagrangian-square-compass}
\end{equation}
Here $K^{'}$ is in principle renormalized differently than $K$ and can be small. The Lagrangian \eqref{eq:Lagrangian-square-compass} is formally similar to that of an AFM. We obtain for the dispersion:
\begin{equation}
	\omega = \sqrt{2JK}Sk.
\end{equation}
This is unsurprisingly the dispersion Eq.~\eqref{eq:dispersion-square-compass} in the large $K$ limit, as the $\rho$-gap becomes large and only the $\phi$-field is retained as the pseudo-Goldstone mode. 

The gradient-dependent interaction in Eq.~\eqref{eq:effective-Lagrangian-square-compass} is just Eq.~\eqref{eq:perturbation-cubic-compass-1} which we have considered in the main text with $K^{'}$ in place of $K$. The free propagator has explicitly the form \eqref{eq:propagator-I}. Thus we obtain immediately the type-I relation Eq.~\eqref{eq:PG-gap-I} for the pseudo-Goldstone gap. 

\section*{References}

\bibliographystyle{unsrt} 

\begin{thebibliography}{10}
	
	\bibitem{Villain1980}
	Jacques Villain, R.~Bidaux, J.-P Carton, and Robert Conte.
	\newblock Order as an effect of disorder.
	\newblock {\em http://dx.doi.org/10.1051/jphys:0198000410110126300}, 41, 11
	1980.
	
	\bibitem{Shender1982}
	E.F. Shender.
	\newblock Antiferromagnetic garnets with fluctuationally interacting
	sublattice.
	\newblock {\em Sov. Phys. JETP}, 56:178, 1982.
	\newblock [Zh. Eksp. Teor. Fiz. {\bf 83}, 326 (1982)].
	
	\bibitem{Henley1989}
	Christopher~L. Henley.
	\newblock Ordering due to disorder in a frustrated vector antiferromagnet.
	\newblock {\em Phys. Rev. Lett.}, 62:2056--2059, Apr 1989.
	
	\bibitem{Champion2003}
	J.~D.~M. Champion, M.~J. Harris, P.~C.~W. Holdsworth, A.~S. Wills,
	G.~Balakrishnan, S.~T. Bramwell, E.~\ifmmode \check{C}\else
	\v{C}\fi{}i\ifmmode~\check{z}\else \v{z}\fi{}m\'ar, T.~Fennell, J.~S.
	Gardner, J.~Lago, D.~F. McMorrow, M.~Orend\'a\ifmmode~\check{c}\else
	\v{c}\fi{}, A.~Orend\'a\ifmmode~\check{c}\else \v{c}\fi{}ov\'a, D.~McK. Paul,
	R.~I. Smith, M.~T.~F. Telling, and A.~Wildes.
	\newblock ${\mathrm{er}}_{2}{\mathrm{ti}}_{2}{\mathrm{o}}_{7}:$ evidence of
	quantum order by disorder in a frustrated antiferromagnet.
	\newblock {\em Phys. Rev. B}, 68:020401, Jul 2003.
	
	\bibitem{Zhitomirsky2012}
	M.~E. Zhitomirsky, M.~V. Gvozdikova, P.~C.~W. Holdsworth, and R.~Moessner.
	\newblock Quantum order by disorder and accidental soft mode in
	${\mathrm{er}}_{2}{\mathrm{ti}}_{2}{\mathbf{o}}_{7}$.
	\newblock {\em Phys. Rev. Lett.}, 109:077204, Aug 2012.
	
	\bibitem{Savary2012}
	Lucile Savary, Kate~A. Ross, Bruce~D. Gaulin, Jacob P.~C. Ruff, and Leon
	Balents.
	\newblock Order by quantum disorder in
	${\mathrm{er}}_{2}{\mathrm{ti}}_{2}{\mathbf{o}}_{7}$.
	\newblock {\em Phys. Rev. Lett.}, 109:167201, Oct 2012.
	
	\bibitem{chandra1990ising}
	P.~Chandra, P.~Coleman, and A.~I. Larkin.
	\newblock Ising transition in frustrated heisenberg models.
	\newblock {\em Phys. Rev. Lett.}, 64:88--91, Jan 1990.
	
	\bibitem{Danu2016}
	Bimla Danu, Gautam Nambiar, and R.~Ganesh.
	\newblock Extended degeneracy and order by disorder in the square lattice
	${J}_{1}\ensuremath{-}{J}_{2}\ensuremath{-}{J}_{3}$ model.
	\newblock {\em Phys. Rev. B}, 94:094438, Sep 2016.
	
	\bibitem{chubukov1992triangular}
	Andrey~V. Chubukov and Th. Jolicoeur.
	\newblock Order-from-disorder phenomena in heisenberg antiferromagnets on a
	triangular lattice.
	\newblock {\em Phys. Rev. B}, 46:11137--11140, Nov 1992.
	
	\bibitem{chubukov1992kagome}
	Andrey Chubukov.
	\newblock Order from disorder in a kagom\'e antiferromagnet.
	\newblock {\em Phys. Rev. Lett.}, 69:832--835, Aug 1992.
	
	\bibitem{Reimers1993}
	Jan~N. Reimers and A.~J. Berlinsky.
	\newblock Order by disorder in the classical heisenberg kagom\'e
	antiferromagnet.
	\newblock {\em Phys. Rev. B}, 48:9539--9554, Oct 1993.
	
	\bibitem{Henley1994}
	Christopher~L. Henley.
	\newblock Selection by quantum fluctuations of dipolar order in a diamond
	lattice.
	\newblock {\em Phys. Rev. Lett.}, 73:2788--2788, Nov 1994.
	
	\bibitem{Baskaran2008}
	G.~Baskaran, Diptiman Sen, and R.~Shankar.
	\newblock Spin-$s$ kitaev model: Classical ground states, order from disorder,
	and exact correlation functions.
	\newblock {\em Phys. Rev. B}, 78:115116, Sep 2008.
	
	\bibitem{Bramwell1994}
	S.~T. Bramwell, M.~J.~P. Gingras, and J.~N. Reimers.
	\newblock Order by disorder in an anisotropic pyrochlore lattice
	antiferromagnet.
	\newblock {\em J. Appl. Phys}, 75:5523--5525, May 1994.
	
	\bibitem{McClarty2014}
	Paul~A. McClarty, Pawel Stasiak, and Michel J.~P. Gingras.
	\newblock Order-by-disorder in the $xy$ pyrochlore antiferromagnet.
	\newblock {\em Phys. Rev. B}, 89:024425, Jan 2014.
	
	\bibitem{Rau2018}
	Jeffrey~G. Rau, Paul~A. McClarty, and Roderich Moessner.
	\newblock Pseudo-goldstone gaps and order-by-quantum disorder in frustrated
	magnets.
	\newblock {\em Phys. Rev. Lett.}, 121:237201, Dec 2018.
	
	\bibitem{Khatua2023}
	Subhankar Khatua, Michel J.~P. Gingras, and Jeffrey~G. Rau.
	\newblock Pseudo-goldstone modes and dynamical gap generation from order by
	thermal disorder.
	\newblock {\em Phys. Rev. Lett.}, 130:266702, Jun 2023.
	
	\bibitem{Tsvelik2005}
	Alexei~M. Tsvelik.
	\newblock {\em {Quantum field theory in condensed matter physics}}.
	\newblock Cambridge Univ. Press, Cambridge, UK, 2 2005.
	
	\bibitem{Auerbach2012}
	A.~Auerbach.
	\newblock {\em Interacting {Electrons} and {Quantum} {Magnetism}}.
	\newblock Graduate {Texts} in {Contemporary} {Physics}. Springer New York,
	2012.
	
	\bibitem{Polyakov1975}
	Alexander~M. Polyakov.
	\newblock {Interaction of Goldstone Particles in Two-Dimensions. Applications
		to Ferromagnets and Massive Yang-Mills Fields}.
	\newblock {\em Phys. Lett. B}, 59:79--81, 1975.
	
	\bibitem{abrikosov2012methods}
	A.A. Abrikosov, L.P. Gorkov, I.E. Dzyaloshinski, and R.A. \~Silverman.
	\newblock {\em Methods of {Quantum} {Field} {Theory} in {Statistical}
		{Physics}}.
	\newblock Dover {Books} on {Physics}. Dover Publications, 2012.
	
\end{thebibliography}

\end{document}